\documentclass[%
aps,
preprint,
showpacs,
preprintnumbers,
 amsmath,
 amssymb,
 prl,
]{revtex4-1}

\usepackage{bm}
\usepackage{dcolumn}
\usepackage{graphicx}

\begin{document}

\title{Extreme case of Faraday effect: magnetic splitting of ultrashort laser pulses in plasmas}

\author{Suming Weng$^{1,2}$}\email{wengsuming@gmail.com}
\author{Qian Zhao$^{1,2}$}
\author{Zhengming Sheng$^{1,2,3}$}\email{z.sheng@strath.ac.uk}%
\author{Wei Yu$^{4}$}
\author{Shixia Luan$^{4}$}
\author{Min Chen$^{1,2}$}
\author{Lule Yu$^{1,2}$}
\author{Masakatsu Murakami$^{5}$}
\author{Warren B. Mori$^{6}$}
\author{Jie Zhang$^{1,2}$}

\affiliation{$^{1}$ Key Laboratory for Laser Plasmas, School of Physics and Astronomy, Shanghai Jiao Tong University, Shanghai 200240, China}%
\affiliation{$^{2}$ Collaborative Innovation Center of IFSA, Shanghai Jiao Tong University, Shanghai 200240, China}%
\affiliation{$^{3}$ SUPA, Department of Physics, University of Strathclyde, Glasgow G4 0NG, UK}
\affiliation{$^{4}$ Shanghai Institute of Optics and Fine Mechanics, Chinese Academy of Sciences, Shanghai 201800, China}%
\affiliation{$^{5}$ Institute of Laser Engineering, Osaka University, Osaka 565-0871, Japan}
\affiliation{$^{6}$ Department of Physics and Astronomy, University of California, Los Angeles, California 90095, USA}

\begin{abstract}
The Faraday effect, caused by a magnetic-field-induced change in the optical properties, takes place in a vast variety of systems from a single atomic layer of graphenes to huge galaxies.
Currently, it plays a pivot role in many applications such as the manipulation of light and the probing of magnetic fields and material's properties.
Basically, this effect causes a polarization rotation of light during its propagation along the magnetic field in a medium.
Here, we report an extreme case of the Faraday effect where a linearly polarized ultrashort laser pulse splits in time into two circularly polarized pulses of opposite handedness during its propagation in a highly magnetized plasma.
This offers a new degree of freedom for manipulating ultrashort and ultrahigh power laser pulses.
Together with technologies of ultra-strong magnetic fields, it may pave the way for novel optical devices, such as magnetized plasma polarizers.
Besides, it may offer a powerful means to measure strong magnetic fields in laser-produced plasmas.
\end{abstract}



\maketitle

\section{Introduction}

As the hallmark of magneto-optics, the Faraday effect or Faraday rotation observed in 1846 was the first experimental evidence of the electromagnetic wave nature of light \cite{Faraday}.
Importantly, it provides an ingenious method for manipulating light, and becomes the basic principle underlying the operation of a number of magneto-optical devices \cite{Saleh,Liu}.
In principle, the Faraday rotation is caused by magneto-chiral dichroism of left-circularly polarized (LCP) and right-circularly polarized (RCP) electromagnetic waves propagating at differential speeds in magnetized materials.
Since the magneto-chiral dichroism in most materials are very weak, considerable Faraday rotation generally happens only after a long propagation distance. This severely limits the miniaturization and integration of magneto-optical devices. Therefore, there has been a growing interest in the search for enhanced Faraday rotation.
As a collection of charged particles, a dense plasmas responds strongly to electromagnetic waves and thus often gives rise to a strong Faraday rotation under the influence of a magnetic field \cite{Chen}.
Furthermore, the plasma optical devices are particularly suitable for the fast manipulation of ultrashort high-power laser pulses due to their ultrahigh damage threshold \cite{Mourou,Thaury,YuLL,Trines}.

\begin{figure}[htbp]
\centering
\fbox{\includegraphics[width=0.75\linewidth]{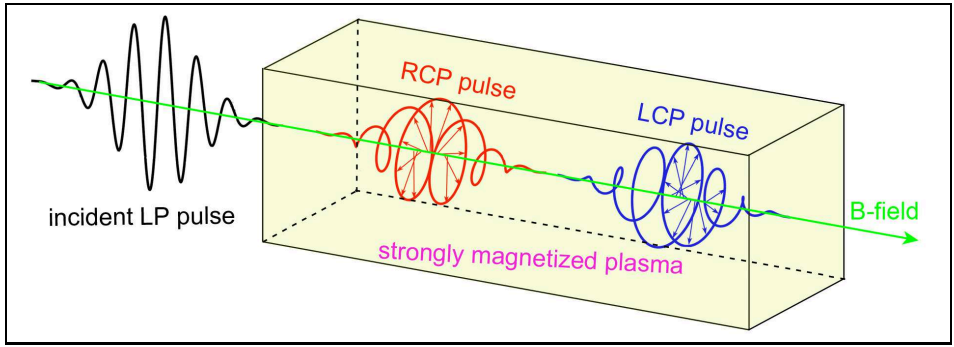}}
\caption{Sketch of the magnetic splitting of an ultrashort LP laser pulse, which is incident along the magnetic field $\textbf{B}$ into plasma. The incident LP pulse will split into RCP and LCP sub-pulses due to their differential group velocities. The RCP sub-pulse follows the LCP sub-pulse in time. Here the electric field vector of the RCP pulse at a fixed position rotates clockwise in time as viewed along the wave vector of the laser pulse, and vice versa for the LCP pulse.}
\label{figSchematic}
\end{figure}

In this work, we report an extreme case of the Faraday effect in which not only the polarization direction but also the polarization state of ultrashort laser pulses can be completely changed in strongly magnetized plasmas with magnetic fields $B \ge 50$ tesla.
The underlying physics is that a linearly polarized (LP) laser pulse can be considered as the superposition of a RCP sub-pulse and a LCP sub-pulse. While the eigen electromagnetic waves propagating along the magnetic field in plasmas are the RCP and LCP waves, which have differential group velocities as well as differential phase velocities.
Therefore, under appropriate conditions a LP laser pulse will split into a RCP sub-pulse and a LCP sub-pulse as shown in Fig. \ref{figSchematic}.

\section{Theory}

We first provide a set of formulas to describe the propagation of electromagnetic waves in magnetized plasmas.
The electromagnetic wave propagation along the magnetic field in a plasma is mainly governed by the dispersion relation \cite{Chen}
\begin{equation}
\frac{c^2k^2}{\omega^2}  = 1- \frac{\omega_p^2}{ \omega^2  (1 \pm \omega_c/\omega)},
\end{equation}
where $\pm$ are respectively for the LCP ($+$) and RCP ($-$) waves, $\omega$ and $k$ are the wave's angular frequency and wavenumber, the plasma frequency $\omega_p\equiv (n_e e^2/\epsilon_0 m_e)^{1/2}$ is defined by the plasma density, and the electron cyclotron frequency $\omega_c\equiv eB/m_e$ is proportional to the magnetic field strength $B$.
From the dispersion relation, one can easily get the phase velocities $v_{p} = [1-\omega_p^2/(\omega^2 \pm \omega \omega_c)]^{-1/2} c$ for the LCP ($+$) and RCP ($-$) waves \cite{Chen}.
The differential phase velocities will induce a rotation of the polarization plane of a LP wave, since it can be considered as the sum of a RCP wave and a LCP wave.
In the limit of low plasma density ($\omega_p \ll \omega $) and small magnetic field ($\omega_c \ll \omega $), the Faraday rotation angle can be estimated as $\Delta \phi \simeq  \texttt{RM} \lambda^2 $,
where $\texttt{RM}=e^3\int n_e(x) B(x) dx/8\pi^2\epsilon_0 m_e^2 c^3$ is the so-called rotation measure in astronomy \cite{Klein,Brandenburg}, and $\lambda$ is the wavelength.
This is the scenario of the familiar Faraday rotation, in which the rotation angle is proportional to the magnitude of magnetic field.
However, this linear Faraday effect can only be applied for a relatively small magnetic field with a low plasma density.
As long as $\omega_c/\omega$ approaches $(1-\omega_p^2/\omega^2)$, the phase velocity for the RCP wave will quickly become infinite. Therefore, the RCP wave cannot propagate in a strong magnetized plasma if $(1-\omega_p^2/\omega^2) \leq \omega_c/\omega \leq 1$.
But the propagation of the RCP wave becomes possible again in the whistler-mode region ($\omega_c/\omega > 1$). In latter case,  the RCP wave can even penetrate into an overcritical density plasma but accompanied by a strong heating of the plasma \cite{YangAPL}.
For the sake of simplicity, we will not discuss the wave propagation in the whistler-mode region here.
Anyhow, one can conclude from the above analysis that the Faraday rotation angle is no longer linearly proportional to the magnitude of the magnetic field if the latter is strong enough, i.e., one enters a nonlinear regime of the Faraday effect.

\begin{figure}[htbp]
\centering
\fbox{\includegraphics[width=0.75\linewidth]{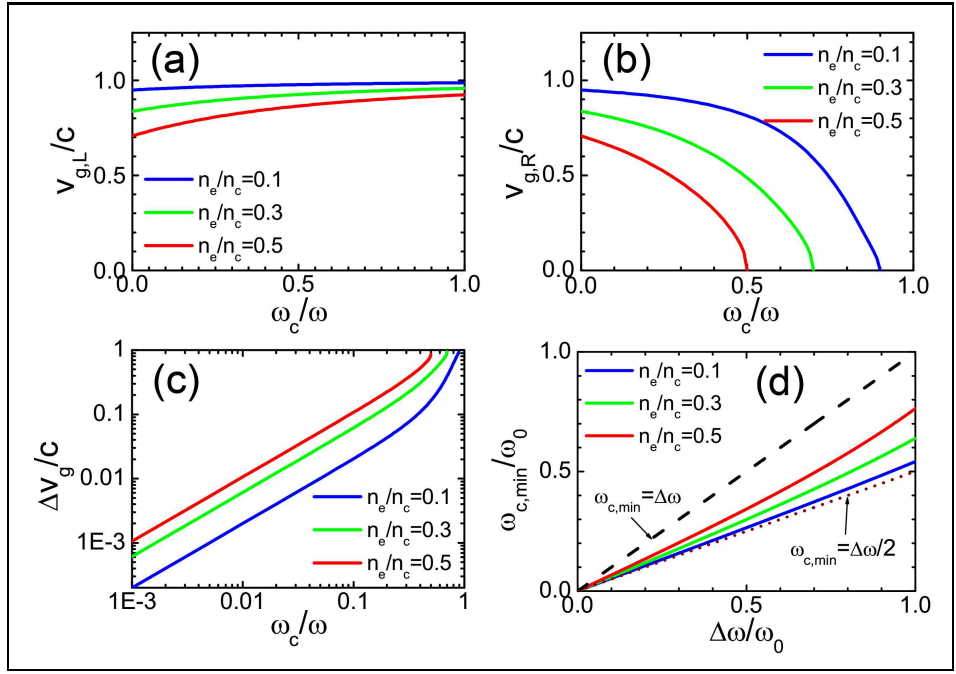}}
\caption{(a,b) Group velocities from Eqs. (\ref{groupL}) and (\ref{groupR}) for the LCP and RCP waves, respectively;
(c) the difference in the group velocities, (d) the minimum field ($\omega_{c,\min}$) required for an obvious magnetic splitting as a function of the  frequency spread ($\Delta\omega$) of the pulse.}
\label{figAnalysis}
\end{figure}

In addition to the differential phase velocities, more importantly, we notice that the group velocities are also different for the LCP and RCP waves in a magnetized plasma. From the dispersion relation, one can deduce the group velocities
\begin{align}
\frac{v_{g,\texttt{L}}}{c} = \left(1-\frac{\omega_p^2/\omega^2}{1 + \omega_c/\omega}\right)^{1/2} \left[1-\frac{\omega_c \omega_p^2/ \omega^3}{2(1+\omega_c/\omega)^2} \right]^{-1},  \label{groupL} \\
\frac{v_{g,\texttt{R}}}{c} = \left(1-\frac{\omega_p^2/\omega^2}{1 - \omega_c/\omega}\right)^{1/2} \left[1+\frac{\omega_c \omega_p^2/ \omega^3}{2(1-\omega_c/\omega)^2} \right]^{-1}, \label{groupR}
\end{align}
for the LCP and RCP waves, respectively.
As shown in Fig. \ref{figAnalysis}, the former increases with the magnetic field, while the latter behaves in the opposite way.
So for a LP short laser pulse, its LCP and RCP components will gradually split apart. Assuming the pulse initially has a duration $t_p$, the time delay between the peaks of LCP and RCP sub-pulses ($\Delta v_gt/ v_{g,R}$) will be larger than $t_p$ after
\begin{equation}
t_s = \frac{v_{g,R}}{\Delta v_g} t_p, \label{tSplit}
\end{equation}
where $\Delta v_g=v_{g,\texttt{L}}-v_{g,\texttt{R}}$ is the difference in the group velocities. Figure \ref{figAnalysis}(c) indicates that the stronger the magnetic field and the higher the plasma density, the larger the difference in the group velocities will be.
If the magnetic field is small enough ($\omega_c \ll \omega$) and the plasma density is low enough ($\omega_p \ll \omega$), we can get
\begin{equation}
\frac{\Delta v_g}{c} \simeq 2 \frac{ n_e }{n_c} \frac{ \omega_c}{ \omega}, \label{deltaVg}
\end{equation}
where $n_c\equiv \epsilon_0 m_e \omega^2/e^2$ is the critical plasma density.

For an ultrashort laser pulse, however, its frequency spread must be taken into account. For instance,
$\Delta\omega/\omega_0 \geq 0.441\tau/t_p$ holds for a Gaussian pulse \cite{Koechner}, where $\tau=2\pi/\omega_0$ is the laser wave period, $\omega_0$, $\Delta \omega$ and $t_p$ are the center frequency, FWHM frequency spread, and FWHM duration of the pulse, respectively.
So that the group velocities are not constant, and the pulse temporal broadening due to dispersion must be considered.
Consequently, the magnetic splitting of the pulse is observable only under the condition
\begin{equation}
v_{g,R}|_{\omega=\omega_0+\Delta\omega/2} < v_{g,L}|_{\omega=\omega_0-\Delta\omega/2}.  \label{omegaMin}
\end{equation}
Otherwise, the dispersive broadening will dominate over the magnetic splitting.
The above inequality prescribes a lower limit of the magnetic field ($B \ge B_{\min}=m_e\omega_{c,\min}/e$).
Under the same assumptions for Eq. (\ref{deltaVg}), we can get $\omega_c \ge \omega_{c,\min} \simeq \Delta\omega/2$, which is in good agreement with the numerical solution at $n_e=0.1n_c$ in Fig. \ref{figAnalysis}(d).
With a relatively higher plasma density such as $n_e=0.5n_c$, however, the required $\omega_{c,\min}$ increases very fast with an increasing $\Delta\omega$.
From the numerical solutions, we find that
\begin{equation}
\omega_c \ge \omega_{c,\min}=\Delta\omega
\end{equation}
is a sufficient condition for the inequality (\ref{omegaMin}) if $n_e/n_c \le 0.5$ as shown in Fig. \ref{figAnalysis}(d).
That is to say, the pulse splitting will be quicker than the dispersive broadening if the electron cyclotron frequency ($\omega_c$) in the magnetic field is larger than the frequency spread ($\Delta \omega$) of the laser pulse. The latter is inversely proportional to the pulse duration.
This implies that the shorter the pulse duration is, the stronger the magnetic field is required to split the pulse.
While Eq. (\ref{tSplit}) implies that the longer the pulse duration is, the thicker the required magnetized plasma has to be.
These two aspects grimly prescribe that the magnetic splitting of a laser pulse can be clearly observed only if the pulse duration is modest and the magnetic field is strong enough.

With the invention of novel laser techniques such as chirped-pulse-amplification \cite{Strickland}, it becomes mature to generate laser pulses as short as femtosecond (fs). On the other hand, the magnets of 20 tesla become commercially available, and the magnetic fields above 100 tesla are recorded in some laboratories\cite{Report}. In particular, the interaction of high-power laser pulses with matters can generate kilo-tesla level magnetic fields \cite{Fujioka,Wagner}. Such kilo-tesla level magnetic fields are not only of fundamental interests, but also show prospects of various applications \cite{Eliezer,HoraLPB,Miley}. The breathtaking advances in the pulsed laser and the high magnetic field sciences conspire to provide the good opportunity to achieve the magnetic splitting of an ultrashort laser pulse.

\section{Simulation}

To verify the magnetic splitting of short laser pulses, we perform a series of particle-in-cell (PIC) simulations using the code OSIRIS \cite{Fonseca}.
In simulations, laser pulses are incident along the magnetic field into semi-infinite plasmas at $x\geq0$.
The initial LP pulses are polarized along the $z$-axis with $\lambda=1\mu\texttt{m}$.
For reference, the pulse peaks are all assumed to arrive at the vacuum-plasma interface (x=0) at t=0.
The moving-window technique is employed with a simulation box moving along the $x$-axis at the speed of light in vacuum. The simulation box is set large enough to contain the laser pulse in the whole process of each simulation.
In 1D simulations, the sizes of the simulation boxes range from $500\lambda$ to $35000\lambda$, the spatial and temporal resolutions are $\Delta x=\lambda/16$ and  $\Delta t \simeq \Delta x/c$, each cell has 16 macro-particles, and the electron density $n_e=0.5n_c$.
In 3D simulation, the simulation box has a size of $210\lambda \times 24000 \lambda \times 24000 \lambda$. The spatial resolutions are $\Delta x=\lambda/16$ and $\Delta y=\Delta z=100 \lambda$, the temporal resolution is $\Delta t \simeq \Delta x/c$, each cell has 4 macro-particles, and the electron density $n_e=0.1n_c$.

\begin{figure}[htbp]
\centering
\fbox{\includegraphics[width=0.75\linewidth]{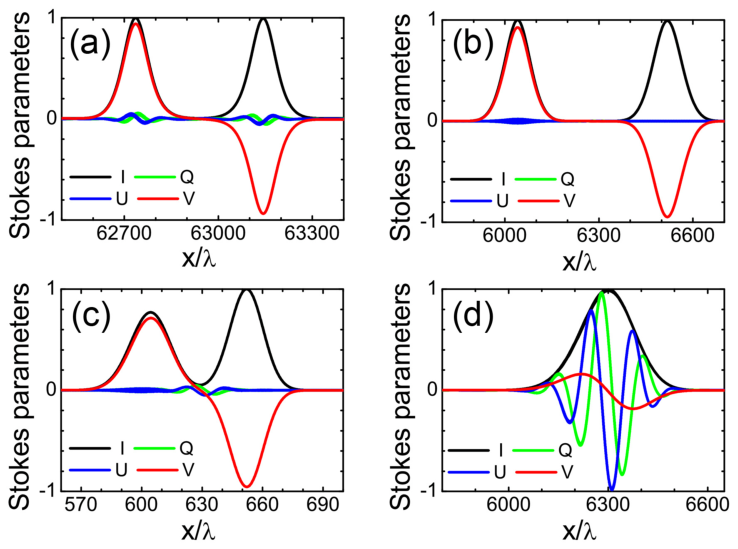}}
\caption{Stokes parameters from 1D PIC simulations with varying laser pulse duration $t_p$ and magnetic field $B$ (a) at $t=300$ ps with $t_p=500$ fs ($\Delta\omega/\omega_0\simeq0.0029$) and B=50 tesla ($\omega_c/\omega_0\simeq0.005$); (b) at $t=30$ ps with $t_p=500$ fs and B=500 tesla; (c) at $t=3$ ps with $t_p=50$ fs, B=500 tesla; and (d) at $t=30$ ps with $t_p=50$ fs, B=50 tesla. The Stoke parameter I denotes the intensity regardless of polarization, Q and U describe the state of linear polarizations, while V represents the circular polarization \cite{Tinbergen}. All parameters are normalized to the instantaneous peak intensity $I_{\max}$. Here the laser intensity is low enough (the dimensionless amplitude $a\equiv|e\textbf{E}/\omega m_e c|=0.01$), so that the nonlinear effects \cite{GibbonBook,HoraBook} due to the laser field itself can be ignored here.
}
\label{fig1D}
\end{figure}

Figure \ref{fig1D} compares 1D simulation results with varying laser pulse duration $t_p$ and magnetic field $B$.
In Fig. \ref{fig1D}(a), the magnetic splitting condition ($\omega_{c}>\Delta\omega$) holds well with $t_p=500$ fs and $B=50$ tesla.
Consequently, the initial pulse splits into two discrete sub-pulses at $t=300$ picosecond (ps).
The first sub-pulse peaked at $x\simeq63145\lambda$ is LCP since its Stokes parameter $V<0$, while the second sub-pulse peaked at $x\simeq62735\lambda$ is RCP with $V>0$.
The degrees of circular polarization exceed $94\%$ for both the LCP and RCP sub-pulses.
Simulation shows that the difference in the group velocities for these two sub-pulses is about 0.0046c, which is in rough agreement with the prediction 0.0050c by Eq. (\ref{deltaVg}).
In Fig. \ref{fig1D}(b), the difference in the group velocities is increased roughly by an order of magnitude with a 500 tesla magnetic field.
Consequently, the LCP and RCP sub-pulses are clearly separated at a much earlier time t=30 ps.
By such a 500 tesla magnetic field, we find that the laser pulses with much shorter durations such as 50 fs can also be separated, although each sub-pulse is a little longer than the initial pulse due to dispersion as shown in Fig. \ref{fig1D}(c).
However, a 50 fs laser pulse cannot be separated by a 50 tesla magnetic field since the pulse frequency spread $\Delta\omega \simeq 0.029\omega > \omega_c \simeq 0.005 \omega$ in this relatively weak magnetic field.
As illustrated in Fig. \ref{fig1D}(d), at t=30 ps the pulse duration has been stretched to about 600 fs, which is an order of magnitude longer than the estimated time delay between the RCP and LCP sub-pulses. This confirms that the dispersive broadening will dominate over the magnetic splitting of the pulse if $\Delta\omega>\omega_{c}$.

\begin{figure}[htbp]
\centering
\fbox{\includegraphics[width=0.5\linewidth]{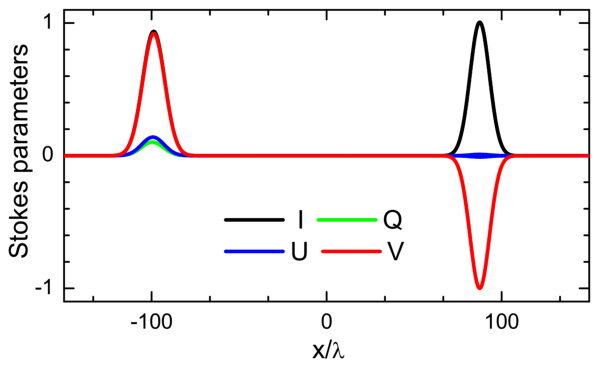}}
\caption{Stokes parameters at $t=333$ fs with an extremely strong magnetic field B=6000 tesla ($\omega_c/\omega_0\simeq0.6$). Other parameters are the same as those in Fig. \ref{fig1D}(c).
}
\label{figRCP}
\end{figure}

Figure \ref{figRCP} displays the simulation result with an extremely strong magnetic field B=6000 tesla ($\omega_c/\omega_0\simeq0.6$). In this case, it becomes impossible for the RCP wave to propagate into the magnetized plasma since $\omega_c/\omega > (1-\omega_p^2/\omega^2)$.
Figure \ref{figRCP} illustrates that the incident LP pulse has been separated into two sub-pulse as well.
However, here only the LCP sub-pulse (peaked at $x\simeq 88 \lambda$ with $V<0$) can propagate into the magnetized plasma. The RCP sub-pulse (peaked at $x\simeq -100 \lambda$ with $V>0$) is completely reflected and propagates backward.

\begin{figure}[htbp]
\centering
\fbox{\includegraphics[width=0.75\linewidth]{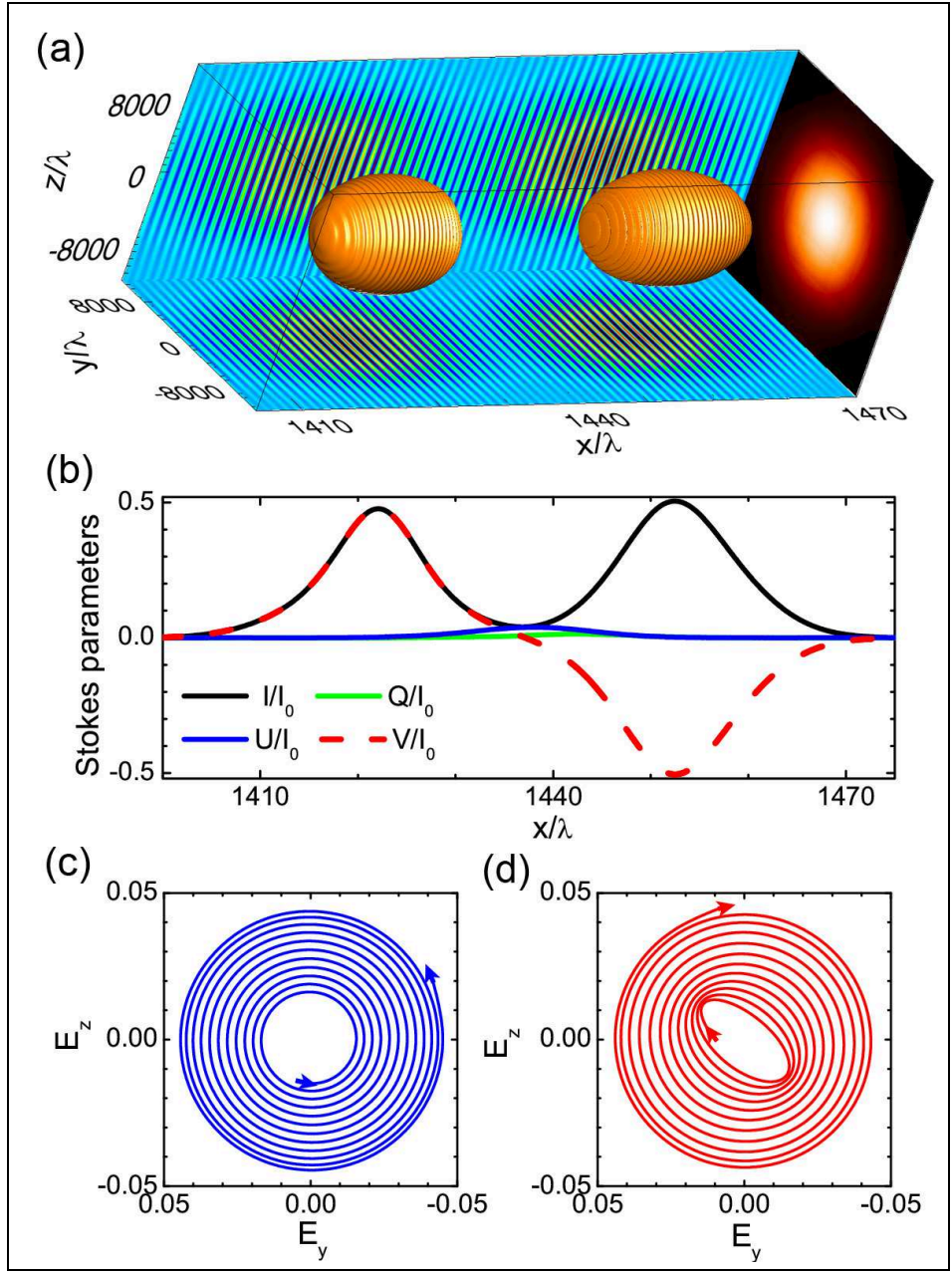}}
\caption{(a) Isosurface of intensity $I=I_0/4$ at $t=5$ ps (yellow ellipsoids). $E_y$ and $E_z$ cross sections at $z=0$ are given on the rear and the bottom of the box, respectively, while the right side displays the transversal distribution of the intensity.
(b) The distributions of the Stokes parameters I, Q, U and V on the $x$-axis, where all parameters are normalized to the initial peak intensity $I_0$.
(c,d) The time evolution of the endpoint of the electric-field vector $\textbf{E}$ in the y-z plane in the time intervals (c) $54\lambda<(ct-x)<66\lambda$ and (d) $85\lambda<(ct-x)<97\lambda$; the arrows indicate that the electric-field vectors rotate anticlockwise and clockwise, respectively, as viewed along  $\textbf{B}$ . Here $E_y$ and $E_z$ are normalized to $m_e \omega c /e$.
} \label{fig3D}
\end{figure}

The magnetic splitting of a 50 fs laser pulse is also verified by a 3D simulation as displayed in Fig. \ref{fig3D}(a), where the isosurface of intensity $I=I_0/4$ appears as two separate ellipsoids at t=5 ps (see Supplementary Movie for the whole splitting process).
At the early stage in the Supplementary Movie, a conventional Faraday rotation as large as many cycles is also evidenced by the quick variations in the $E_y$ and $E_z$ components of the electric field.
Here a relatively lower plasma density $n_e=0.1n_c$ is used to alleviate nonlinear effects \cite{GibbonBook}, and a stronger magnetic field B=1000 tesla is employed in order to save the computation time.
The laser intensity $I_0=1.37\times10^{16}$ W/cm$^2$ ($a_0=0.1$), and the peak power is 10 petawatt (PW) with a waist $r_0 \simeq 6800\lambda$.
The intensity distribution on the $x$-axis in Fig. \ref{fig3D}(b) suggests that two sub-pulses have the FWHM durations $\approx 47$ fs and the peak intensities $I_{\max}\approx I_0/2$ as expected according to energy conservation.
Since the laser intensity now is already weakly relativistic, each sub-pulse is a little shorter than the initial pulse due to the self-compression of intense laser pulses in plasmas \cite{Shorokhov}.
The first sub-pulse centered at $x\simeq 1452\lambda$ has a group velocity $v_{g,L}\simeq 0.955c$, while the second one at $x\simeq 1422 \lambda$ has $v_{g,R}\simeq0.935c$.
They are in good quantitative agreement with the predictions by Eqs. (\ref{groupL}) and (\ref{groupR}), respectively.
And the difference between these two group velocities is approximate to the estimation by Eq. (\ref{deltaVg}).

Regardless of the temporal splitting of the pulse, the transversal distribution of the laser intensity is keeping as a Gaussian function as shown in Fig. \ref{fig3D}(a).
Furthermore, Fig. \ref{fig3D}(a) illustrates that the y-component of the electric field $E_y$ at t=5 ps becomes as strong as the z-component $E_z$, although the pulse is initially polarized along the $z$-axis only. Figure \ref{fig3D}(c) illuminates that the endpoint of the electric-field vector rotates anti-clockwise as viewed along $\textbf{B}$ in the time interval $54 \lambda<(ct-x)<66 \lambda $. This time interval corresponds to the rising stage of the first sub-pulse, and the electric-field vector at its falling stage also rotates anti-clockwise and thus is omitted here. Therefore, we are convinced that the first sub-pulse is a LCP pulse and hence it propagates faster. Conversely, Fig. \ref{fig3D}(d) confirms that the endpoint of the electric-field vector rotates clockwise during $ 85 \lambda<(ct-x)<97 \lambda$ and the second sub-pulse is a RCP pulse.

\section{Discussion}

In comparison with the familiar Faraday rotation, the extreme case of the Faraday effect reported above offers a new degree of freedom to manipulate ultrashort high power laser pulses.
Therefore, it may form the basis of a new type of novel optical devices, such as magnetized plasma polarizers.
Since the laser gain of amplifiers and the loss of resonators such as Brewster plate are usually polarization dependent, the laser emissions are often linearly polarized \cite{Koechner}.
To get a circularly polarized pulse, a quarter-wave plate is employed in common \cite{Saleh}.
For a high power laser pulse, however, the quarter-wave plate suffers from the problem of optically induced damage \cite{Koechner}.
The state-of-the-art laser facilities under construction will provide a peak power as high as 10 PW \cite{Papadopoulos}, where the diameter of the quarter-wave plate should be larger than a few decimetres to avoid the laser-induced damage.
To the best of our knowledge, it is extremely challenging to manufacture such a large-diameter quarter-wave plate.
Fortunately, one may realize a novel type of magnetized plasma polarizer for such high power lasers based on the above extremely strong Faraday effect.
Thanks to the ultrahigh damage threshold of plasmas, this magnetized plasma polarizer is nearly free from laser-induced damage. It is worthwhile to notice that in the above 3D simulation the laser pulse already has a peak power of 10 PW, and this pulse has been converted into circularly polarized sub-pulses by a magnetized plasma on the centimeter scale (a waist of 0.68 cm).
The resultant high-power circularly polarized pulses are particularly attractive to the laser-driven ion acceleration\cite{Daido,Macchi}, the optical control of mesoscopic objects\cite{Arita}, and the ultrahigh acceleration of plasma blocks for fusion ignition\cite{Eliezer,HoraLPB,Miley,Weng}.

Although the magnetized plasma polarizer is nearly free from laser-induced damage, it also has its own limitations due to nonlinear effects in intense laser-plasma interactions \cite{GibbonBook,HoraBook}.
Above all, the laser pulse may collapse at a distance $\sim z_R (P/P_c)^{-1/2}$ if its power exceeds the critical power for relativistic self-focusing ($P_c \simeq 17.5 n_c/n_e $ GW) \cite{GibbonBook}, where $z_R=\pi r_0^2/\lambda$ is the Rayleigh length. Therefore, the distance for the magnetic splitting ($\sim ct_s$) must be shorter than $z_R (P/P_c)^{-1/2}$. Using Eqs. (\ref{tSplit}) and (\ref{deltaVg}), we can get
\begin{equation}
r_0^2 \frac{\omega_c}{\omega} (\frac{n_e}{n_c})^{1/2} > \frac{\lambda  t_p v_{g,R} }{2\pi} [\frac{P}{17.5 \texttt{ GW}}]^{1/2}.
\end{equation}
This prescribes a lower limit for the pulse waist $r_0$.
Assuming $\omega_c=0.01\omega$ ($B\simeq 100$ T) and $n_e/n_c=0.1$, we find that a waist $r_0 > 1700\lambda$ is required for the magnetic splitting of a 500 fs 10 PW laser pulse.
Setting $a_0=0.1$ for a 10 PW laser pulse, we will have a pulse waist $r_0 \simeq  6800\lambda$ that satisfies the above requirement well.
With $r_0 \simeq 6800\lambda$, the Rayleigh length $z_R \simeq 1.44 \times 10^{8} \lambda$.
On the other hand,  such a large waist and a long Rayleigh length are also crucial in postponing the self-modulational instability, which is due to laser-driven plasma wakefield and becomes significant at the time scale of laser self-focusing \cite{GibbonBook,Antonsen,Esarey}.
Secondly, besides the relativistic self-focusing, the laser pulse could also be focused by a transversely inhomogeneous plasma with $dn(r)/dr>0$ or defocused with $dn(r)/dr<0$.
Analogous to the geometric optics picture of self-focusing in Ref. \cite{GibbonBook}, we can get $\Delta v_p/c \sim \Delta n_e/2n_c$, where $\Delta v_p$ ($\Delta n_e$) is the difference between the phase velocities (plasma densities) at the center and at the edge of the pulse.
Then the focusing (or defocusing) angle of the laser pulse is given by $\alpha \simeq  \sqrt{\Delta v_p/c}  = \sqrt{\Delta n_e/2n_c}$.
Further, the condition $\alpha < r_0/ct_s$ should be satisfied in order to split the laser pulse before it is focused (or defocused).
Combining this condition with Eqs. (\ref{tSplit}) and (\ref{deltaVg}), we can get
\begin{equation}
\frac{\Delta n_e}{n_e} < 8 \frac{n_e (r_0 \omega_c )^2} {  n_c (c t_p \omega)^2}. \label{Dne}
\end{equation}
Under the conditions $\omega_c=0.01\omega$ and $n_e/n_c=0.1$,  it is required that $\Delta n_e/n_e<16.2\%$ for the magnetic splitting of a 500 fs 10 PW laser pulse with a waist $r_0 \sim 6800 \lambda$.
Similarly, we can get the difference in phase velocity due to the transverse inhomogeneity of magnetic field as $\Delta v_p/c \sim n_e \Delta \omega_c/2n_c \omega$, where $\Delta \omega_c$ is the difference between the magnetic fields at the center and at the edge of the pulse.
In the case of $n_e=0.1n_c$ and $\omega_c/\omega=0.01$, we find that $\Delta v_p/c \sim 0.0005 \Delta \omega_c/\omega_c $ will be very small. Consequently, the focusing or defocusing effect due to the transverse inhomogeneity of magnetic field could be negligible in this case. However, a magnetic field inhomogeneity less than a few tens of percentages would be of great benefit to the quality of the resultant LCP and RCP sub-pulses.
The magnetic splitting of laser pulses should not be sensitive to the longitudinal inhomogeneity of plasma density or magnetic field. For a  longitudinally inhomogeneous plasma or/and magnetic field, we find that the distance between the peaks of LCP and RCP sub-pulses is approximate to $\int \Delta v_g dt  \simeq \int \Delta v_g dx/c \propto \int n_e(x) B(x) dx$, and the magnetic splitting emerges if this distance is larger than $c t_p$.
Thirdly, if gaseous targets are used, one should also take into account the nonlinear effects due to ionization and Kerr nonlinearity.
The former could induce a defocusing effect since usually more electrons are produced via ionization on the laser axis.
While the latter could induce a self-focusing effect since the higher intensity at the pulse center leads to a larger refractive index.
It is worthwhile to point out that these nonlinear effects sometimes may counteract each other. For instance, a plasma channel as long as a few kilometres in the atmosphere could be created if the Kerr effect balances the diffraction and the ionization-induced defocusing \cite{GibbonBook,Kasparian}.

Due to the nonlinear effects discussed above, the laser pulse will lose energy as it propagates in a plasma even if the collisional damping is ignored.
From simulations, we find that it is crucial to set $a_0 \ll 1$ and $n_e \ll n_c$ in order to reduce the collisionless losses.
So we use $a_0=0.1$ and $n_e=0.1n_c$ in the 3D simulation shown in Fig. \ref{fig3D}.
Then about 95.092\% laser energy can be preserved in the LCP (48.069\%) and RCP (47.023\%) pulses. In particular, only about $0.062\%$ laser energy is lost after $t=100$ fs when the pulse propagates inside the plasma, other $4.846\%$ laser energy is lost near the vacuum-plasma interface before $t=100$ fs.
Therefore, one can expect that the collisionless losses can be controlled at a level of a few percentages with a much longer propagation distance when a relatively weaker magnetic field ($\sim 100$ T) and a longer laser pulse ($\sim 500$ fs) are used.
The collisional losses, which are not treated in our PIC simulations, can be estimated as $K_{ib}=1-\exp{(-\kappa_{ib} L)}$ \cite{EliezerBook}, where $L \sim c t_s$ is the distance required for the magnetic splitting and $\kappa_{ib} \simeq  \nu_{ei} (n_e/n_c)^2(1-n_e/n_c)^{-1/2}/c$ is the spatial damping rate by inverse bremsstrahlung.
At high laser intensities, e.g. $I > 10^{15}$ W/cm$^2$, the electron-ion collision frequency should be modified as $\nu_{ei}\simeq Z_i e^4 n_e \ln \Lambda / (4 \pi \epsilon_0^2 m_e^2 v_{eff}^3)$ \cite{GibbonBook, EliezerBook, WengPRE}, where $Z_i$ is the ionization state, $\ln \Lambda$ is the Coulomb logarithm, and the effective electron thermal velocity $v_{eff} = (v_{te}^2 + v_{os}^2)^{1/2} \simeq a_0 c$ is defined by the electron thermal velocity $v_{te}$ and the electron oscillatory velocity $v_{os} \simeq a_0c$ in the laser field.
Assuming $\omega_c=0.01\omega$, $n_e=0.1n_c$, and $Z_i \ln \Lambda \simeq 10$, we can get $K_{ib} \simeq 7.2\%$ for a 500 fs laser pulse with $a_0=0.1$.
With a decreasing plasma density, we find that both the collisionless losses and the collisional losses can be reduced.
With a decreasing laser intensity, however, the collisional losses will increase although the collisionless losses can be reduced. A moderate laser intensity $\sim 10^{16}$ W/cm$^2$ ($a_0 \sim 0.1$) may be appropriate to keep both the collisonal and collisionless losses at a tolerable level.

Besides the applications as optical devices, this extremely strong Faraday effect may be applied to measure ultra-strong magnetic fields. Although the Faraday rotation is widely used in the measurement of magnetic fields, it essentially has three limitations.
Firstly, the magnetic field should be small enough ($\omega_c\ll\omega$) to guarantee its linear relation with the Faraday rotation angle.
Secondly, there may be $n\times180^o$ ambiguity of the Faraday rotation angle.
Thirdly, the exact information of the initial polarization direction is required.
In laser-produced plasmas with strong magnetic fields ($B \sim 1000T$) \cite{Wagner,Fujioka}, sometimes it may be difficult to meet all above requirements together.
In these scenarios, however, the probe pulse may split into two circularly polarized pulses due to the extremely strong Faraday effect if the plasma thickness $>100\lambda n_c/n_e$ (the corresponding areal density $\rho R > 10^{-4}$ g/cm$^2$).
Then the magnetic field could be estimated from the time delay between two resultant circularly polarized pulses.
Therefore, this extremely strong Faraday effect could be a powerful alternative to the conventional Faraday rotation in the measurement of ultra-strong magnetic fields in plasmas.
Such strongly magnetized plasmas may be encountered in magnetically assisted fast ignition \cite{WangMAFI}, which is advantageous in depositing the laser energy into the core of fuel target in inertial confinement fusion.

It is worth pointing out that the higher the plasma density is, the more obvious this extremely strong Faraday effect is. This is because the light is slowed down more obviously and the difference in the group velocities is larger at a higher plasma density.
We notice that the temporal splitting of laser pulses can also be achieved in other slow-light medium such as atomic vapors\cite{Grischkowsky}, although in which the pulse duration is usually longer than nanosecond.
In contrast to a bandwidth of gigahertz for the tunable pulse with atomic vapors\cite{Camacho}, the femtosecond laser pulses with terahertz (THz) bandwidths can be manipulated by the magnetized plasmas.
In principle, this extremely strong Faraday effect can be applied to manipulate electromagnetic radiations from radio waves to gamma rays for multiple potential applications \cite{Gibbon,Teubner,Wang}.
However, this effect is observable only when $\omega_c/\omega$ is not too small, which presents a practical limit for experiments at high wave frequency.
While for a THz radiation, magnetic fields on the order of tesla are already enough to achieve this effect.

\section{Conclusion}

In summary, an extreme case of the Faraday effect has been found in magnetized plasmas due to its remarkable chiral dichroism. With this, the magnetic splitting of a LP short laser pulse into a LCP pulse and a RCP pulse can be realized. This opens the way for advanced applications, such as a magnetized plasma polarizer. The latter could allow the generation of circularly polarized laser pulses as high-power as 10 PW in the up-to-date laser facilities. Moreover, this eliminates some limits in the Faraday rotation for the measurement of magnetic fields, thus offering a way to measure ultra-high magnetic fields, broadly existing in objects in the universe and laser-matter interactions in the laboratories.

\section*{Funding Information}
National Basic Research Program of China (2013CBA01504); National Natural Science Foundation of China (NSFC) (11129503, 11374210, 11405108, 11421064, 11675108); National 1000 Youth Talent Project of China; Leverhulme Trust.

\section*{Acknowledgments}
The authors thank L. J. Qian, J. Q. Zhu, J. Fuchs, S. Chen, Y. T. Li, Z. Zhang, T. Sano, H. C. Wu, X. H. Yuan, Y. P. Chen, G. Q. Xie, and L. L. Zhao for fruitful discussions. Simulations have been carried out at the Pi cluster of Shanghai Jiao Tong University.

\bigskip \noindent See \href{link}{Supplement 1} for supporting content.



\bibliography{apssamp}

\end{document}